\documentclass[journal=jacsat,manuscript=article]{achemso}
\usepackage[version=3]{mhchem} 
\usepackage[dvipsnames]{xcolor}

\author{Greta Segantini}
\affiliation{Department of Quantum Matter Physics, University of Geneva, 24 Quai Ernest-Ansermet, CH-1211 Geneva 4, Switzerland.}
\author{Chih-Ying Hsu}
\affiliation{Electron Spectrometry and Microscopy Laboratory (LSME), Institute of Physics (IPHYS), Ecole Polytechnique Fédérale de Lausanne (EPFL), CH-1015 Lausanne, Switzerland.}
\alsoaffiliation{Department of Quantum Matter Physics, University of Geneva, 24 Quai Ernest-Ansermet, CH-1211 Geneva 4, Switzerland.}
\author{Carl Willem Rischau}
\affiliation{Department of Quantum Matter Physics, University of Geneva, 24 Quai Ernest-Ansermet, CH-1211 Geneva 4, Switzerland.}
\author{Patrick Blah}
\affiliation{Kavli Institute of Nanoscience, Delft University of Technology, 2628 CJ Delft, The Netherlands.}
\author{Mattias Matthiesen}
\affiliation{Kavli Institute of Nanoscience, Delft University of Technology, 2628 CJ Delft, The Netherlands.}
\author{Stefano Gariglio}
\affiliation{Department of Quantum Matter Physics, University of Geneva, 24 Quai Ernest-Ansermet, CH-1211 Geneva 4, Switzerland.}
\author{Jean-Marc Triscone}
\affiliation{Department of Quantum Matter Physics, University of Geneva, 24 Quai Ernest-Ansermet, CH-1211 Geneva 4, Switzerland.}
\author{Duncan T.L. Alexander}
\affiliation{Electron Spectrometry and Microscopy Laboratory (LSME), Institute of Physics (IPHYS), Ecole Polytechnique Fédérale de Lausanne (EPFL), CH-1015 Lausanne, Switzerland.}
\author{Andrea D. Caviglia}
\affiliation{Department of Quantum Matter Physics, University of Geneva, 24 Quai Ernest-Ansermet, CH-1211 Geneva 4, Switzerland.}
\email{greta.segantini@unige.ch}

\title[An \textsf{achemso} demo]
  {Electron-Beam Writing of Atomic-Scale Reconstructions at Oxide Interfaces}


\begin{document}




\begin{abstract}
The epitaxial growth of complex oxides enables the production of high-quality films, yet substrate choice is restricted to certain symmetry and lattice parameters, thereby limiting the technological applications of epitaxial oxides. In comparison, the development of free-standing oxide membranes gives opportunities to create novel heterostructures by non-epitaxial stacking of membranes, opening new possibilities for materials design. Here, we introduce a method for writing, with atomic precision, ionically bonded crystalline material across the gap between an oxide membrane and a carrier substrate. The process involves a thermal pre-treatment, followed by localized exposure to the raster scan of a scanning transmission electron microscopy (STEM) beam. STEM imaging and electron energy-loss spectroscopy show that we achieve atomically sharp interface reconstructions between a 30 nm-thick SrTiO${_3}$ membrane and a niobium-doped SrTiO${_3}$(001)-oriented carrier substrate. These findings indicate new strategies for fabricating synthetic heterostructures with novel structural and electronic properties. 
\end{abstract}

\section{Main text}
Complex oxides exhibit a broad spectrum of functionalities, including ferroelectricity, ferromagnetism, and high-temperature superconductivity \cite{Chambers2010,koster2015epitaxial}. In recent years, significant attention has been directed toward their potential applications across various technological domains \cite{Cen2009, Li2020, Liu2022}.
Epitaxial growth enables the fabrication of high-quality oxide films, providing an ideal platform for investigating their physical properties at the atomic level. 
Moreover, interface engineering of epitaxially-grown oxide layers has led to the discovery of intriguing interface phenomena \cite{Caviglia2008,Reyren2007,Zubko2011}. 
However, the epitaxial relationship between thin film and substrate imposes limitations on the application of stimuli to the oxides, such as strain, and confines the substrate selection to those meeting specific symmetry and lattice spacing requirements. This constraint extends to the integration of epitaxial oxides into the complementary metal-oxide-semiconductor (CMOS)-based industry, which at present remains a significant challenge. 
Inspired by the isolation of 2D materials, such as graphene and transition-metal dichalchogenides, a promising way to overcome intrinsic limitations of epitaxial oxides is to detach them from their growth substrate. 
Among the strategies explored, the chemical lift-off approach has gained considerable interest \cite{Lu2016, Sanchez-Santolino2023, Wang2020}. In this approach, epitaxially-grown sacrificial layers are dissolved using suitable etchants, thereby releasing oxide layers as membranes that can be transferred and stacked, free of epitaxial restrictions. As well as opening a pathway for creating new heterostructures, this approach potentially leads to the integration of epitaxial oxides into CMOS-based technology.
Literature reports have demonstrated the remarkable response of oxide membranes to strain \cite{Dong2019,Elangovan2020,Harbola2021}, and the ability to control the twist angle of stacked membranes to create and manipulate moiré patterns \cite{Li2022,Sanchez-Santolino2023}. These systems further hold promise for applications in nanoelectronics, including non-volatile memories, sensors, and flexible electronics \cite{Pesquera2020, Dong2020, Lee2022}. While research on oxide membranes has yielded innovative results, the ability to create a strong chemical bond between the membrane and a carrier substrate (or second membrane) onto which it is transferred remains relatively unexplored.

\begin{figure}[H]
\centering
  \includegraphics[width=\textwidth]{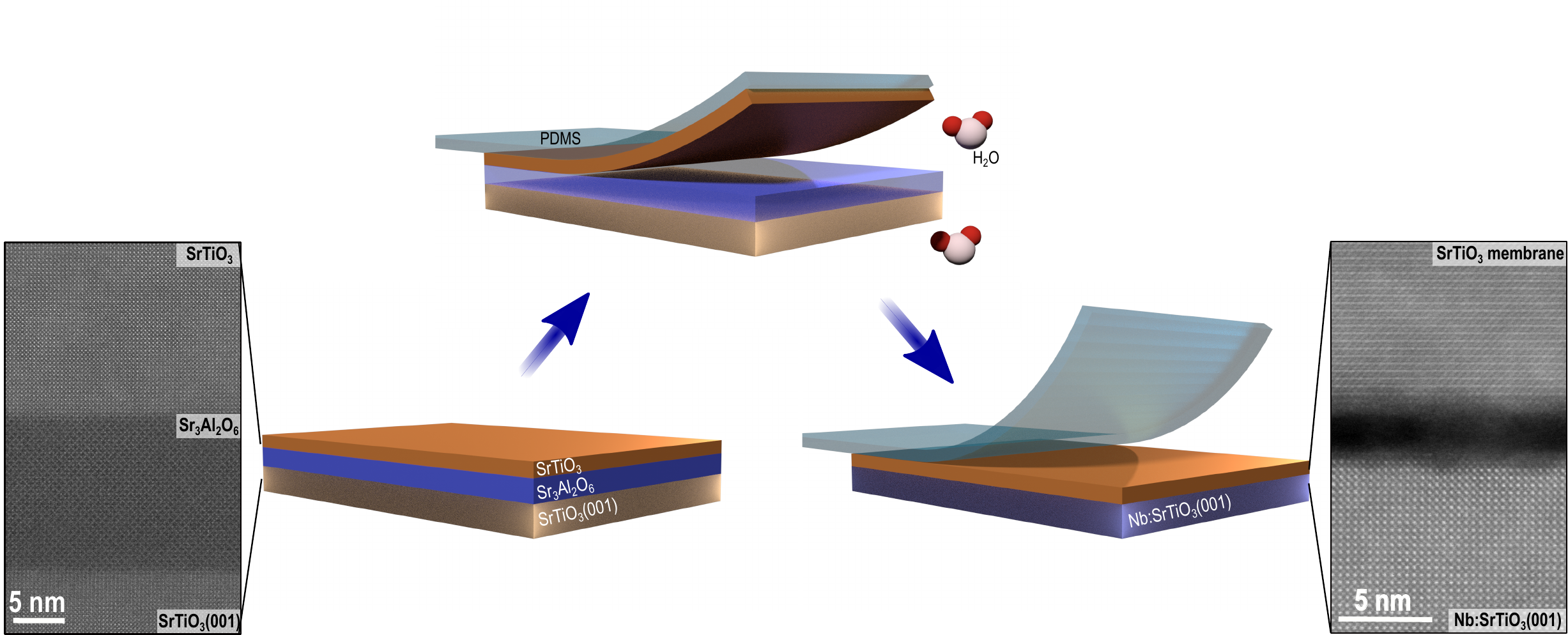}
  \caption{Schematic of the membrane fabrication process: The Sr$_{3}$Al$_{2}$O$_{6}$ sacrificial layer and the SrTiO${_3}$ membrane were synthesized using PLD. Subsequently, a PDMS sheet was applied to the surface of the SrTiO${_3}$ layer, and the entire structure was immersed in deionized water. Following the dissolution of the Sr$_{3}$Al$_{2}$O$_{6}$ layer, the resulting SrTiO${_3}$ membrane was transferred onto the Nb:SrTiO${_3}$(001) substrate. HAADF cross-section images of the heterostructure SrTiO${_3}$(001)/Sr$_{3}$Al$_{2}$O$_{6}$/SrTiO${_3}$ before lift-off and of the SrTiO${_3}$ membrane transferred onto the Nb:SrTiO${_3}$(001) substrate are shown on the left and on the right, respectively. Here, an interface gap of $\sim$2 nm between the SrTiO${_3}$ membrane and the Nb:SrTiO${_3}$(001) substrate is distinguishable.}
\label{fig:membrane_fabrication}
\end{figure}

Here, we report on the control of interfacial ionic bonds between a 30 nm-thick SrTiO${_3}$ membrane and a niobium-doped SrTiO${_3}$(001)-oriented (Nb:SrTiO${_3}$) carrier substrate.
The SrTiO${_3}$ membranes were fabricated by epitaxial growth of a 15 nm-thick Sr$_{3}$Al$_{2}$O$_{6}$ sacrificial layer followed by a 30 nm-thick SrTiO${_3}$ layer on a SrTiO${_3}$(001)-oriented substrate using pulsed-laser deposition (PLD). As schematically illustrated in \textbf{Figure \ref{fig:membrane_fabrication}}, a strip of polydimethylsiloxane (PDMS) was applied to cover the entire surface of the SrTiO${_3}$ layer for the lift-off process, and the structure was immersed in deionized water at room temperature to dissolve the Sr$_{3}$Al$_{2}$O$_{6}$ layer. The resulting SrTiO${_3}$ membrane was then transferred onto a Nb:SrTiO${_3}$(001)-oriented non-terminated substrate, and the PDMS removed. The Supporting Information (SI) provides details on the transfer procedure, along with other experimental parameters. 
Figure \ref{fig:membrane_fabrication} also shows high-angle annular dark-field (HAADF) STEM images of sample cross-sections: the SrTiO${_3}$(001)/Sr$_{3}$Al$_{2}$O$_{6}$/SrTiO${_3}$ heterostructure before lift-off (left), and the SrTiO${_3}$ membrane after transfer (right). Combined with the X-ray diffraction patterns of SI Figure \ref{fig:XRD}, these images show that the good crystalline quality of the SrTiO${_3}$ membrane is preserved during transfer. We note that the initially-flat layer does sometimes acquires some low amplitude modulations after transfer, as measured using atomic force microscopy (SI Figure \ref{fig:AFM}).
In the following, three samples are studied. One is as-transferred (Sample 0), while the other two underwent an additional thermal annealing step at atmospheric pressure for 1 h, at temperatures of 550 $^\circ$C (Sample 550) or 750 $^\circ$C (Sample 750).

\begin{figure}[H]
\centering
  \includegraphics[width=\textwidth]{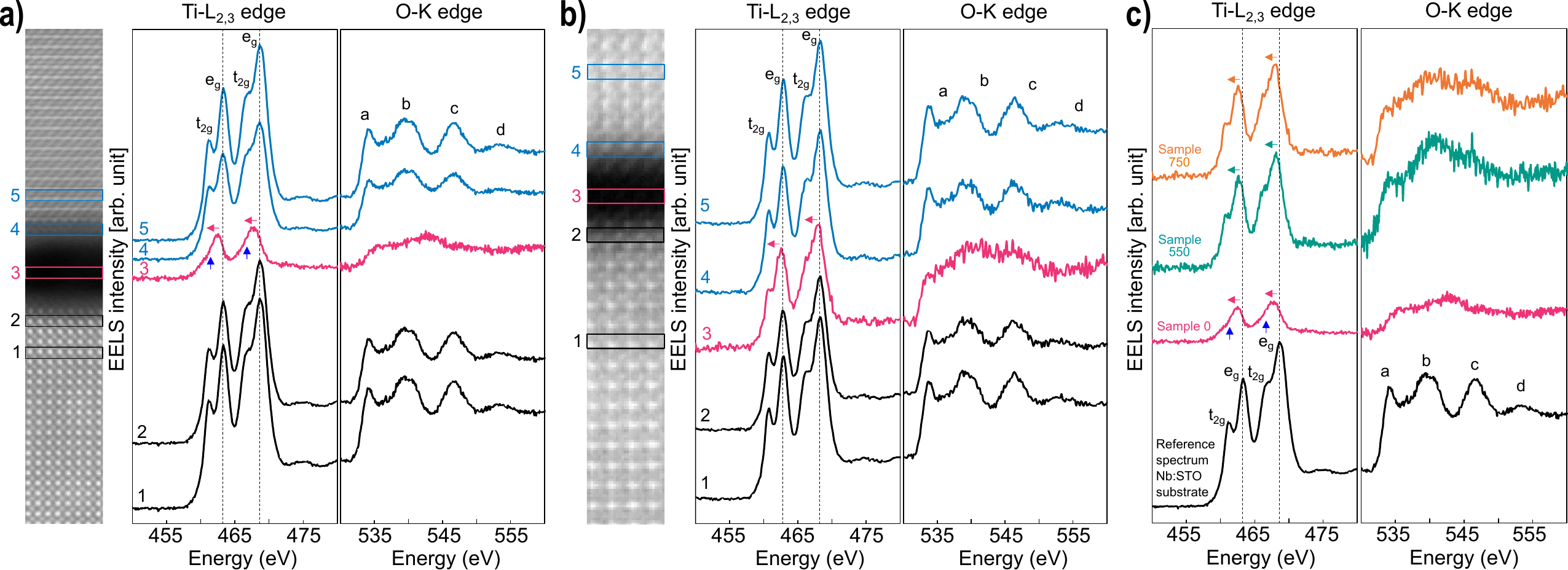}
  \caption{Effect of annealing of the SrTiO${_3}$ membrane on Nb:SrTiO${_3}$(001) substrate. a) STEM-EELS analysis of Sample 0. From left to right: HAADF cross-section image of Nb:SrTiO${_3}$(001) substrate/SrTiO${_3}$ membrane, background-subtracted Ti-$L_{2,3}$ and O-$K$ edges extracted from \#1 Nb:SrTiO${_3}$(001) substrate, \#2 Nb:SrTiO${_3}$(001) substrate near the bottom interface, \#3 center of the gap, \#4 SrTiO${_3}$ membrane near the top interface, and \#5 SrTiO${_3}$ membrane. The Ti-$L_{2,3}$ and O-$K$ edges in spectrum \#3 reveal a clear change compared to the crystalline SrTiO${_3}$. In particular, the splitting of the Ti $L_{3}$ and $L_{2}$ peaks observed in spectra \#1, \#2, \#4, and \#5, that indicates Ti$^{4+}$ oxidation state, is no longer visible, suggesting a change in Ti valence from $4+$ to $2+$. b) STEM-EELS analysis of Sample 750: spectrum \#3 of Ti-$L_{2,3}$ edge indicates that Ti has shifted towards $4+$ valence; the first fine structure peaks of the O-$K$ edge has also moved to a lower energy compared to gap spectrum a). c) Comparison of Ti-$L_{2,3}$ and O-$K$ edges obtained from Sample 0, Sample 550, and Sample 750 extracted in the center of the gap together with a reference from the Nb:SrTiO${_3}$(001) substrate. The evolution of the Ti-$L_{2,3}$ edges as a function of the annealing temperature demonstrates a clear change in Ti valence state. All the displayed spectra are background-subtracted, equivalently normalized by substrate intensities, and aligned on the energy-loss axis using the O-$K$ edge onset energy (532 eV). Note that, for compactness, the HAADF images are cropped from the full width of the original mapped areas. The EEL spectra are instead integrated from the full map width.} 
\label{fig:annealing_effect}
\end{figure}

First, we examine the effect of annealing on the substrate/membrane system. Outside of any height-modulated membrane regions, the adjacent crystalline surfaces of the Nb:SrTiO${_3}$(001) substrate and SrTiO${_3}$ membrane are relatively smooth and uniformly spaced, with a gap between them that evolves under the thermal annealing (see low magnification cross-section STEM images in SI Figure \ref{fig:ADF_large_area}). In Sample 0, the gap measures $\sim$2 nm in width. The origin of the gap is associated with the presence of contamination species on the substrate surface and on the membrane surface in contact with the Sr$_{3}$Al$_{2}$O$_{6}$ sacrificial layer, stemming from its dissolution in deionized water \cite{Wang2024}. \textbf{Figure \ref{fig:annealing_effect}a} shows a higher magnification image of the interface gap in Sample 0. Given that the intensity of a HAADF image $I \propto Z^{1.6-1.9} $ (average atomic number $Z$) \cite{Hartel1996}, the dark contrast of the gap is attributed to its amorphous, disordered nature and its lower density compared to the crystalline material either side. Energy dispersive X-ray spectroscopy (EDXS) was used to analyse the elements present within the gap. Major elements of strontium, titanium and oxygen were found, together with carbon, a common contaminant from air exposure, and calcium and aluminum, which are residuals from the lift-off process (SI Figure \ref{fig:EDX_sample0}). 
Further, electron energy-loss spectroscopy (EELS) was used to study the electronic/bonding state of the Ti and O going across the gap from the substrate to the membrane. As shown in Figure \ref{fig:annealing_effect}a, the Ti $L$-edge of the crystalline substrate (spectra \#1 \& \#2) presents the signature splitting of Ti $L_{2}$ and $L_{3}$ peaks. This results from spin-orbit coupling, which gives rise to two distinct peaks that are attributed to the $t_{2g}$ and $e_g$ molecular orbitals, characteristic of the Ti$^{4+}$ in octahedral symmetry \cite{de19902, Stemmer2000}. The O $K$-edge in turn presents a series of well-defined peaks (labeled a, b, c, and d) that are characteristic of SrTiO${_3}$ \cite{Stemmer2000}. As expected from its high-quality crystalline nature, spectrum \#5 from the membrane shows equivalent features to the Nb:SrTiO${_3}$(001) substrate. However, spectrum \#3 from the middle of the gap is distinctly different. Owing to the lower density of material, the edge intensities are strongly reduced. Further, no splitting is visible in the Ti $L_{2,3}$ peaks, which are also left-shifted by $\sim$1 eV. 
Equally, the first peak of the $O$ K-edge is shifted to a higher energy. These observations are consistent with a Ti valence in the gap of $\sim$Ti$^{2+}$, and a loss of octahedral coordination with oxygen atoms \cite{Stoyanov2007}. We note that, despite spectra \#2 \& \#4 being extracted close to the interface, they still display the same characteristics as those obtained from the substrate and membrane.

Figure \ref{fig:annealing_effect}b shows the STEM-EELS analysis for Sample 750 (see SI Figure \ref{fig:EELS_550} for Sample 550). The thermal annealing induces a number of changes to the gap. First, its width reduces to $\sim$0.9 nm. At the same time, the normalized intensities of the Ti and O edges within the gap are more than doubled compared to Sample 0. Together, these imply a densification from annealing, without loss of material content. Figure \ref{fig:annealing_effect}b spectrum \#3 from the gap also shows that the ionization edge structures are modified. Both the Ti $L_{2}$ and $L_{3}$ peaks present a discernible splitting, with a reduced left shift, and the first peak of the O $K$-edge shifts towards the edge onset. Both these results suggest that annealing has moved the Ti valence state up from Ti$^{2+}$ towards Ti$^{4+}$.

To profile the spectral evolution from annealing, Figure \ref{fig:annealing_effect}c shows EEL spectra extracted from the interface gaps of samples 0, 550, and 750, together with a reference spectrum from the Nb:SrTiO${_3}$(001) substrate. Each spectrum is normalized in intensity by its dataset's substrate spectrum, and aligned by the onset energy of the O $K$-edge at 532 eV. The figure underscores the onset of splitting and reduced left shift of the Ti $L_{2}$ and $L_{3}$ peaks with thermal annealing. Also, the O $K$-edge transitions towards having features similar to those of the substrate reference spectrum.
Overall, therefore, the STEM-EELS analyses show that the annealing procedure not only reduces gap width between membrane and Nb:SrTiO${_3}$(001) substrate, but also modifies the Ti valence state at the interface from Ti$^{2+}$ towards Ti$^{4+}$.

Finally, we point out the in-plane structural misalignment observed between the membrane and the substrate in the HAADF cross-section images of Figures \ref{fig:annealing_effect}a and \ref{fig:annealing_effect}b, which were both acquired along a reference zone axis of the substrate. 
As the membranes are not in zone axis, their atomic columns cannot be (clearly) distinguished. From tilting the sample stage, this misalignment was quantified to be $\sim$2$^\circ$ for Sample 0 and $\sim$0.7$^\circ$ for Sample 750, and is considered a natural consequence of applying a small twist during the manual transfer process.

\begin{figure}[H]
\centering
  \includegraphics[width=\textwidth]{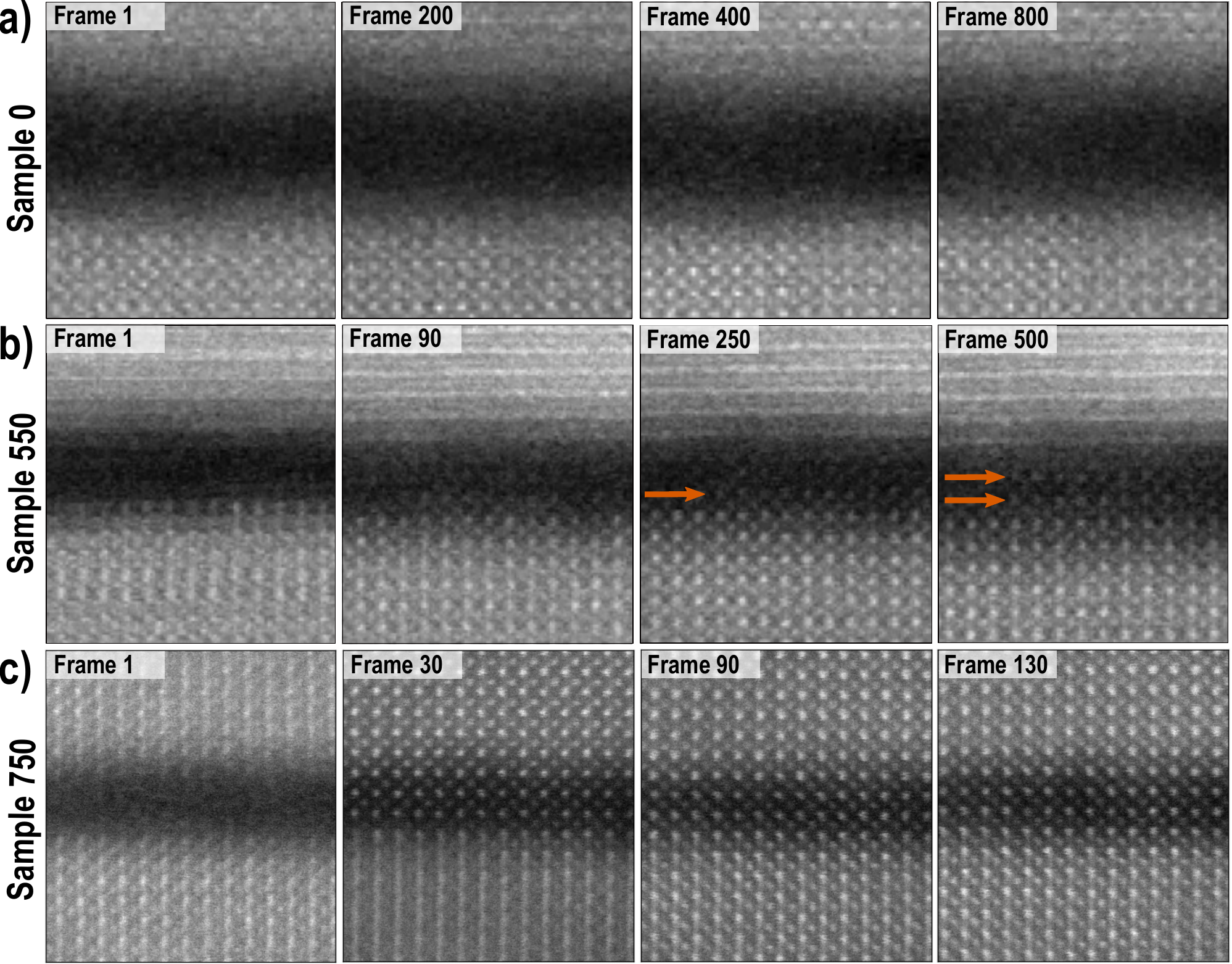}
  \caption{Effect of STEM-EDXS raster scan conditions on Nb:SrTiO${_3}$(001) substrate/SrTiO${_3}$ membrane system, showing the evolution of the interface gap as a function of the acquired number of frames. a) Sample 0: no evident changes are observed after 800 frames. b) Sample 550, ordered atomic structure emerges within the gap after 250 frames. At the final frame 500, the gap is largely filled with crystalline structure. c) Sample 750, crystal structure completely fills the gap after 30 frames. By frame 130, the cation sites of substrate and membrane have also come into alignment.} 
\label{fig:beam_effect}
\end{figure}

In the second part of this study we look at the impact of the STEM electron-beam (``e-beam'') on the substrate/membrane interface. The data presented in the previous section were taken using acquisition conditions that were carefully tuned in order to measure the three samples in their original condition (see Experimental in SI). However, we observed that, when the e-beam flux is above a certain threshold value (discussed below), rastering it across the $\sim$0.9 nm interface gap of an annealed sample leads its structural modification. 
\textbf{Figure \ref{fig:beam_effect}} illustrates this structural evolution. Each row presents frames from HAADF STEM image series of the three samples, that were acquired under ``STEM-EDXS'' conditions (2 \text{$\mu$s} dwell time, 250 pA beam current, multiple frame series). In Figure \ref{fig:beam_effect}a) for Sample 0 no change is observed within the gap, even after 800 frames.  
In contrast, in Figures \ref{fig:beam_effect}b,c for Samples 550 and 750, it is evident that, under the 250 pA e-beam raster scan, crystal structure forms within the gap.
For Sample 550, new atomic structure within the gap first becomes visible after 250 frames, corresponding to a cumulative electron dose of $\sim$3.04 $\times$ 10$^{6}$ e$^-$~\AA$^{-2}$. Seemingly, it propagates from the substrate towards the membrane, as indicated by the orange arrows. By frame 500 (electron dose $\sim$6.08 $\times$ 10$^{6}$ e$^-$~\AA$^{-2}$), crystalline structure bridges the full gap, as marked by two orange arrows.

Figure \ref{fig:beam_effect}c portrays the same analysis for Sample 750, where new crystalline structure has formed across the whole gap after just 30 frames, corresponding to an electron dose of $\sim$2.92 $\times$ 10$^{6}$ e$^-$~\AA$^{-2}$).  
Remarkably, structural transformation continues; by frame 130 the interface region is fully reconstructed, with the image showing that, either side of the crystallized gap, the cation sites of membrane and substrate have themselves come into alignment. SI Figure \ref{fig:haadf_profiles} shows the accompanying atomic resolution EDXS elemental maps, integrated from the full series of mapping frames, where the net count line profiles confirm that the brighter and darker cation rows across the interface region respectively correspond to Sr and Ti.

In the case of Sample 750, the new crystal structure appears to propagate from the membrane towards the substrate, in contrast with Figure \ref{fig:beam_effect}b for Sample 550. After repeated image series acquisitions of different regions under the same EDXS mapping parameters but different sample tilts, we find that these opposing observations are mostly a consequence of the chosen imaging condition. In fact, we conclude that the crystal structure that forms within the gap originates from both sides. However, our observation of structural propagation is sensitive to the alignment of the crystal structure to the incident e-beam; atomic columns are much more distinct when the crystal is aligned very close to a perfect zone axis condition (i.e. a condition with strong electron channeling down the atomic columns). Therefore, when the substrate is better aligned to the incident beam, as in Figure \ref{fig:beam_effect}b, the new structure appears to propagate from the substrate; when instead the membrane is better aligned, it appears to start from the membrane (Figure \ref{fig:beam_effect}c).

\begin{figure}[H]
\centering
  \includegraphics[width=\textwidth]{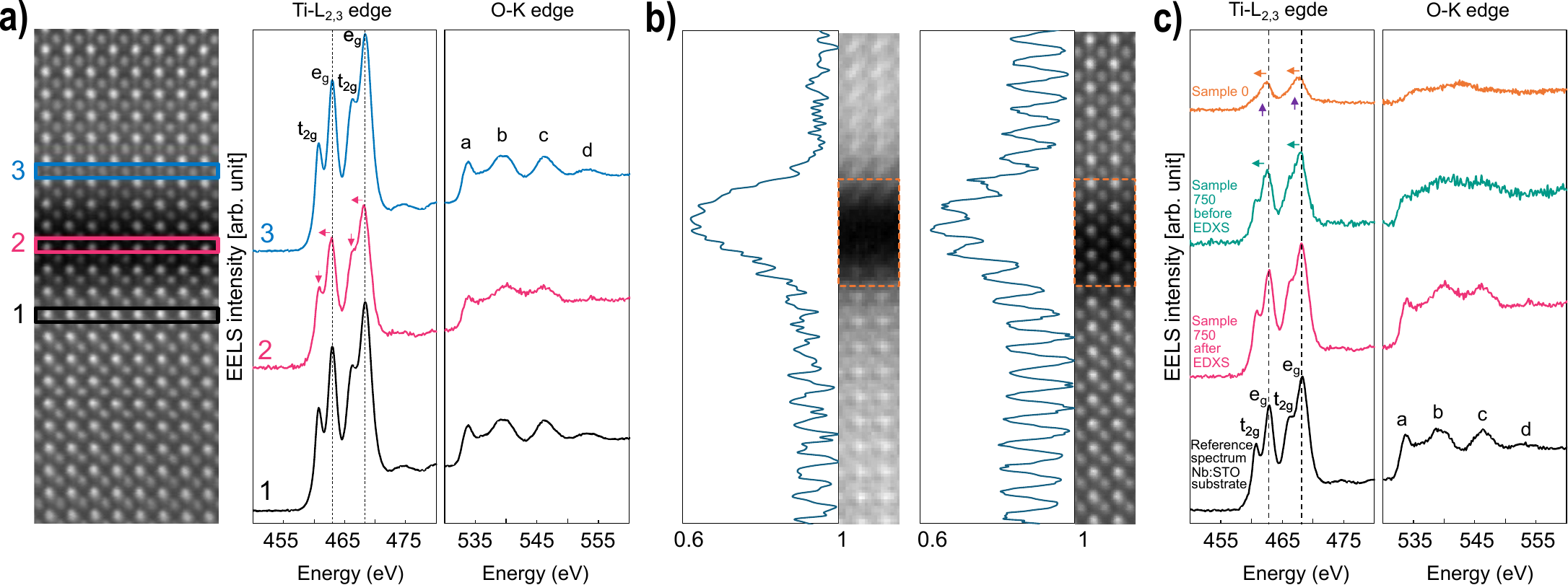}
  \caption{Effect of the EDXS condition e-beam raster scan on Sample 750. a) STEM-EELS analysis after the EDXS raster scanning that is shown in Figure \ref{fig:beam_effect}c. From left to right, the HAADF image, Ti-$L_{2,3}$ edge, and O-$K$ edges acquired from the Nb:SrTiO${_3}$(001) substrate, the gap, and the membrane. The HAADF image depicts clear atomic columns within the gap. EEL spectral features of the Ti-$L_{2,3}$ and O-$K$ edges closely resemble those observed in the Nb:SrTiO${_3}$(001) substrate and membrane. b) Ti-$L$ edge integrated signal from EELS map acquired from the initial ``pristine'' area, and from the same area after the EDXS raster scanning shown in Figure \ref{fig:beam_effect}c. The consistent intensity of the Ti integrated signal indicates no mass loss or gain of Ti atoms within the gap during the structural reorganization to a crystalline structure.The HAADF images are cropped from the full width of the original mapped areas. c) Comparison of EEL spectra from the center of the interface gap for as-transferred, 750 $^\circ$C annealed, and 750 $^\circ$C annealed--EDXS raster scanned, together with a reference spectrum from the Nb:SrTiO${_3}$(001) substrate. Displayed EEL spectra were processed as for Figure \ref{fig:annealing_effect}.} 
\label{fig:comparison_annealing_beam_effect}
\end{figure}

In \textbf{Figure \ref{fig:comparison_annealing_beam_effect}} we study these bridging crystalline structures using EELS, in the case of Sample 750. Figure \ref{fig:comparison_annealing_beam_effect}a shows the Ti-$L$ and O-$K$ edges at the same region where the EDXS scanning of Figure \ref{fig:beam_effect}c was made. Unlike the as-annealed condition of Figure \ref{fig:annealing_effect}b, the Ti-$L_{2,3}$ and O-$K$ edges at the centre of the interface region now closely resemble those of the membrane and Nb:SrTiO${_3}$(001) substrate. This indicates the formation of SrTiO${_3}$ crystal structure, such that ionic bonds have formed across the interface gap. (To help illustrate the evolution in EELS fine structure, SI Figure \ref{fig:projected_edges} presents EEL spectra projected along a line in the out-of-plane direction from before and after the local e-beam irradiation.) 

In Figure \ref{fig:comparison_annealing_beam_effect}b, we consider mass preservation during the e-beam induced restructuring. HAADF images from before and after EDXS scanning show that, after the EDXS scanning, clear atom columns are visible across the gap (marked in orange boxes). Next to the HAADF images, we plot respective line-profiles of the Ti-$L$ edge integrated signal. While in the ``after'' case, the line profile acquires strong modulations corresponding to the new, well-defined atomic planes, the overall Ti signal intensity remains unchanged. In both cases, it shows a drop of $\sim$40\% at the gap center compared to substrate/membrane. This confirms, on the one hand, the less dense nature of the gap compared to the substrate and membrane and, on the other hand, that the crystal structure observed after EDXS scanning derives from reorganization of existing Ti cations into octahedral coordination with oxygen atoms, without incorporating extra Ti cations from elsewhere. 

Figure \ref{fig:comparison_annealing_beam_effect}c provides a summary of spectral progression across sample processing, showing gap EEL spectra for the as-transferred (orange), 750 $^\circ$C-annealed (green), and 750 $^\circ$C-annealed+raster-scanned conditions (pink). Only in the latter case do the Ti-$L_{2,3}$ and O-$K$ edge features align with those of the reference spectrum from the bulk substrate (black), showing a Ti valence state modification from Ti$^{2+}$ to Ti$^{4+}$. With a combination of annealing and e-beam raster scanning, the originally amorphous gap has been bridged by ionic bonds between the membrane and the Nb:SrTiO$_3$(001) substrate, complemented by the octahedral coordination of the residual oxygen atoms.

We now consider the mechanisms behind the e-beam induced writing of atomic structure. The powerful scope for using the STEM analytical probe to create and tailor structures down to the atomic scale is generally established \cite{dyck2019atom}. Further, a recrystallization of ion-beam amorphized SrTiO${_3}$ on SrTiO${_3}$ single crystal under e-beam raster scanning, into a perfect, epitaxial SrTiO${_3}$ crystal lattice, was previously observed by Jesse et \textit{al.} \cite{Jesse2015}. Sample damage or modification by an incident e-beam is typically ascribed to one of two basic possibilities: first, a ballistic interaction of the fast transmitting electron with an atomic nucleus that leads to atomic displacement (knock-on damage); second, the excitation of an atomic electron above the Fermi level, temporarily leaving a hole that may destabilize the atom's bonding, which then leads to a change in atomic bonding and structure (radiolysis) \cite{Egerton2019}.
Jesse et \textit{al.} hypothesized that the crystallization of amorphous SrTiO${_3}$ was promoted by knock-on damage\cite{Jesse2015}. While knock-on often sputters material, we assume that it was instead conceived to locally rearrange atomic species without mass loss. We, however, hypothesize that in our case radiolysis is the more critical factor. In this context, radiolysis electronically excites the atoms in the interface gap. When these atoms recover to a ground state, they go to a new, more stable state as they rearrange to form the observed, ionically bonded crystal structure. Potentially, the ionization of Ti cations directly aids transfer of their charge to the O anions and the correlated formation of strong ionic bonds. In support of our hypothesis, we identify an upper bound of electron flux of $\sim$10$^{10}$ e$^-$~\AA$^{-2}$ s$^{-1}$ that avoids crystallization of the gap region for Sample 750, as exemplified for EDXS in SI Figure \ref{fig:impact_current}. (See SI section E-Beam Flux Effects for details.) This is hard to understand in the context of pure knock-on, where damage is perceived as being permanent and hence proportional to e-beam dose and non-recoverable under any flux. Knock-on damage is also typically associated with mass-loss from sputtering, which we do not observe (Figure \ref{fig:comparison_annealing_beam_effect}b). However, the threshold is consistent with radiolysis, when atoms are allowed to recover to their initial ground state under sufficiently low flux. To emphasize the critical nature of this electron flux, when using a 0.25 \AA~step-size, increasing the e-beam current used for the EELS mapping from 90 pA to 100 pA was enough to induce observable atomic rearrangements within the gap during multiple pass acquisitions. An analogous nature of flux thresholds has been observed for preserving the pristine O sublattice of crystalline samples of cuprates and nickelates \cite{Haruta2018,Mundet2021}. 

As mentioned earlier, continued exposure to the e-beam raster scan not only produces crystalline structure bridging the gap, but can realign adjacent zones of substrate and membrane that are also exposed. In Figure \ref{fig:beam_effect}c this leads to the local ``untwisting'' of membrane and substrate to the same zone axis orientation, producing the high quality lattice spanning across them in Figure \ref{fig:comparison_annealing_beam_effect}a. Since the surrounding bulk lattices of both the substrate and membrane remain unaffected, these local displacements must be accommodated by strong local lattice distortions and/or defect creation. As such distortions are primarily in-plane, it is however difficult to discern them using cross-section imaging. 
Recently, Wang et \textit{al.} interpreted the structural rotation of the first two monolayers of an SrTiO${_3}$ membrane, which had been bonded to a sapphire substrate, by subjecting them to a 1000 $^\circ$C laser-induced, ultra-high vacuum thermal anneal~\cite{Wang2024}.
To give a comparative indication of the possible zone of distortion associated with the e-beam writing of crystalline structure here, SI Figure \ref{fig:focus_window} presents lower magnification HAADF STEM images of Sample 750 before and after a local e-beam raster scan (this time made using a ``focus window'').

After exposure, crystalline lattice 7$-$10 unit cells deep into the substrate or membrane has twisted into a new configuration that differs from the unexposed surrounding area. In this case, the two pre-existing crystals appear to have become less aligned during the reconfiguration. However, in our interpretation we cannot control for the effects of translational displacements or membrane subgrain boundaries that are invisible in the projection of the STEM image. Nevertheless, it is clear that the structurally-affected zone penetrates far from the interface, implying a strong effect of bonding across the gap that is consistent with the formation of ionic bonds.
Such strong effects even extend to another 750 $^\circ$C annealed sample having a larger membrane/substrate misalignment of $\sim$4$^\circ$, which shows both full gap reconstruction and $\sim$3 unit cell substrate realignment after sufficient EDXS raster scanning (SI Figure \ref{figimpact_mistilt}).
 
Finally, we point out that, within the gap, the new crystalline structure can extend a couple of unit cells laterally beyond the region directly impacted by the raster scan (green box in SI Figure \ref{fig:focus_window}b). This suggests that incoherently-scattered secondary electrons, or coherent excitations with longer interaction lengths (plasmons, phonons), may also play a role in the crystallization \cite{Egerton2019,Kisielowski2021}, hinting at a complexity of interactions that needs further investigation to understand fully.

In summary, we demonstrate the local writing of crystalline structure across the interface gap between a 30 nm-thick SrTiO${_3}$ membrane and a Nb:SrTiO${_3}$ (001)-oriented substrate, achieved through two steps. First, a thermal annealing of the bulk sample, that reduces the gap width and shifts the valency of the residual Ti and O atoms in the gap from TiO equivalents to more oxidized species. Second, a STEM e-beam raster scan that induces the Sr, Ti and O atoms in the interface gap to rearrange into ionically bonded crystalline lattice. 
The results indicate that the annealing temperature strongly influences the efficacy of interface reconstruction by the e-beam raster scan, with a slower process observed for the sample annealed at 550 $^\circ$C compared to that annealed at 750 $^\circ$C. 
We note that, in an extra sample that was annealed for 3 h at 750 $^\circ$C, and that contained bumps in the transferred membrane, it was possible to crystallize across gaps of $\sim$4 nm in width. This suggests that the valence evolution from thermal annealing is more critical for the e-beam-induced crystallization to occur than the reduction in gap size.

The method introduced here allows nanometric precision in crystal structure writing. For each annealed condition the extent of structural transformation can be controlled by tuning a combination of electron flux and total dose, with the possibility of completely avoiding reconstruction by staying below threshold flux values. With sufficient flux and dose, structural effects can propagate up to $\sim$10 unit cells deep into the substrate and membrane, creating localized lattice strains.
 
This precise control of interface reconstruction between perovskite oxides therefore represents a powerful tool for discovering new interfacial phenomena, for instance by inducing tailored strain gradients. Further, employing other electron probes, such as scanning electron microscopy (SEM) e-beams or e-beam lithography, could enable the interface reconstruction over large areas within the membrane/substrate system. We note that, since ionization cross-section increases as beam energy decreases, a radiolysis-driven reconstruction is fully compatible with use of SEM. Our approach therefore paves a way for developing synthetic oxide membrane-based heterostructures, with the ability to selectively induce ionic bonding between them.

\begin{acknowledgement}
This work was supported by the Swiss State Secretariat for Education, Research and Innovation (SERI) under contract no. MB22.00071, the Gordon and Betty Moore Foundation (grant no. 332 GBMF10451 to A.D.C.), the European Research Council (ERC), and by the Netherlands Organisation for Scientific Research (NWO/OCW) as part of the VIDI (project 016.Vidi.189.061 to A.D.C.), the ENW-GROOT (project TOPCORE) programmes. We acknowledge the Interdisciplinary Centre for Electron Microscopy (CIME) at EPFL for providing access to their electron microscopy facilities.

\end{acknowledgement}

\bibliography{achemso-demo}

\begin{suppinfo}

\section{Experimental}

\subsection {Sample Preparation}
The 30 nm-thick SrTiO${_3}$ layer was epitaxially grown on a 15 nm-thick Sr$_{3}$Al$_{2}$O$_{3}$ sacrificial layer on a (001)-oriented non-terminated SrTiO${_3}$ substrate. The films were grown using pulsed-laser deposition (PLD) equipped with a KrF excimer laser ($\lambda= 248$ nm). The Sr$_{3}$Al$_{2}$O$_{6}$ layer was deposited on the SrTiO${_3}$(001) substrate using a laser fluency of 2.2 J.cm$^{-2}$, and the SrTiO${_3}$ layer with a laser fluency of 1.6 J.cm$^{-2}$. Both layers were grown at 700 $^\circ$C with an oxygen pressure of $10^{-5}$ mbar. Following the deposition, the sample was cooled down in the same oxygen pressure for 180 minutes. 
The lift-off procedure was performed by applying a polydimethylsiloxane (PDMS) strip to the SrTiO${_3}$ surface, followed by transferring the structure to de-ionized water for approximately 1 h to dissolve the Sr$_{3}$Al$_{2}$O$_{6}$ sacrificial layer. Subsequently, the PDMS with the resulting SrTiO${_3}$ membrane was transferred onto the Nb:SrTiO${_3}$(001) (0.5 wt\%) non-terminated substrate. The transfer process consisted in placing the Nb:SrTiO${_3}$(001) substrate on a hot plate maintained at 80 $^\circ$C. Once the PDMS/membrane system adhered to the substrate, it was pressed down for 90 s.
The annealing procedure was performed in a tubular furnace kept at atmospheric pressure.

\subsection{Atomic Force Microscopy}
The surface morphology analysis was conducted using AFM with a \textit{Digital Instrument} Nanoscope Multimode DI4 with a \textit{Nanonis} controller. 

\subsection{X-Ray Diffraction}
XRD measurements were performed using a \textit{Panalytical X’Pert} diffractometer with Cu K$\alpha$1 radiation (1.54 \AA) equipped with a 2-bounce Ge(220) monochromator and a triple axis detector. 

\subsection{TEM Lamella Preparation}
The TEM lamellae in this study were prepared by focused ion beam milling using a Zeiss NVision 40. The lamellae preparation follows a typical preparation procedure with milling beam energy of 30 keV and currents varying from 1.5 nA down to 80 pA. At the end of milling, each lamella was cleaned using a 5 keV beam with a current of 30 pA. The final thicknesses of the lamellae are estimated to be between approximately 40 - 70 nm according to low-loss EELS analyses.

\subsection{STEM Analyses}
The STEM analyses were carried out on a double aberration-corrected FEI Titan Themis 80-300, using a 300 keV beam energy and 20 mrad probe convergence semi-angle. This tool is equipped with the four quadrant ChemiSTEM SuperX EDXS system, and a GATAN GIF Continuum ERS EEL spectrometer with K3 direct detection camera that was operated in counting mode. EDXS data were acquired using Thermo Fisher Scientific Velox software, from which HAADF image series are exported. The displayed EDXS datasets were recorded using a 250 pA beam current and a 2 \text{$\mu$s} dwell time. However, additional experiments were made using beam currents of 100 and 150 pA in order to investigate electron flux (i.e. dose rate) thresholds for sample modification. 

EELS data were recorded using Gatan DigitalMicrograph 3.5, and with a semi-angle of collection on the spectrometer of 40.8 mrad. The EELS acquisition parameters were rigorously selected to avoid any beam-induced structural modification of the interface gap, as deduced from comparing HAADF images recorded before and after the EELS mapping. These parameters were: singleEELS mode with 0.34 ms dwell time per pixel (minimum possible dwell time for the K3 detector), and the combinations of e-beam current and pixel step-size listed in Table 1 below. These parameters also give near 100\% duty cycle during spectral acquisition, therefore maximising the utilization of the incident electrons. Each map corresponds to 6 integrated frames. Active drift correction was applied between acquisition of each frame, with drift measured using a separate reference region. Table 1 also gives an example condition that showed evidence of structural changes during the dataset acquisition from an annealed sample (100 pA beam current, 0.25 \AA~step-size). This condition could however be safely applied for the data shown in Figure \ref{fig:comparison_annealing_beam_effect}a, since that sample had already been modified by the prior EDXS mapping.

\begin{table}[H]
  \centering
  \caption{EELS acquisition conditions of dataset shown for different samples.}
    \begin{tabular}{lc|cccc}
    Sample & Figure & Dwell time  (ms) & Current (pA) & Step-size (\AA) & \# of frames \\
    \hline
    \hline
    Sample 0 & 2a & 3.4 & 100   & 0.5   & 6 \\
    Sample 550 & S5 & 3.4 & 90    & 0.25  & 6 \\
    Sample 750 & 2b & 3.4 & 90    & 0.5   & 6 \\
    750 after EDXS & 4a & 3.4 & 100   & 0.25  & 6 \\
    \textcolor{red}{Sample modified} & \textcolor{ red}{} & \textcolor{red}{3.4} & \textcolor{red}{100} & \textcolor{red}{0.25} & \textcolor{red}{6} \\
    \hline
    \end{tabular}%
  \label{tab:EELS_condition}%
\end{table}%

In terms of data processing, the EEL spectrum images were denoised using principal component analysis with the standard MSA function available in DigitalMicrograph 3.5. From the denoised datacubes, pixel integrated spectra were extracted, and spectral background removed using a standard power-law fitting. The intensity scales of displayed spectra are normalized by the Nb:SrTiO${_3}$(001) substrate spectrum of each dataset, and calibrated on the energy axis by the O-$K$ edge onset energy (532 eV) in the substrate, to facilitate the comparison between different datasets. We note that deconvolution of multiple scattering was not applied, because the increased electron flux of dualEELS mode was incompatible with avoiding beam-induced structural changes.

\section {E-Beam Flux Effects}

As detailed in Experimental, the effects of different e-beam currents on the annealed SrTiO${_3}$ membrane/Nb:SrTiO${_3}$(001) substrate samples were tested, for both EDXS (2 \text{$\mu$s} dwell time; 100, 150 and 250 pA) and EELS (0.34 ms dwell time; 90 and 100 pA) mapping acquisition conditions, combined with various pixel step-sizes.

In the EDXS case, no structural evolution of the interface gap of Sample 750 was observed when applying the lowest e-beam current of 100 pA, as demonstrated by the series of HAADF image frames in SI Figure \ref{fig:impact_current}. Applying a current of 150 pA instead led to slow structural changes. Under EELS conditions, the equivalent e-beam current limit for avoiding structural modification is 90 pA when using a 0.25 \AA~step-size. Notably, these limiting conditions \textit{both} correspond to an electron flux in the region of $\sim$10$^{10}$ e$^-$~\AA$^{-2}$ s$^{-1}$, even though pixel dwell times were very different.

The EDXS tests at lower beam currents were made on a fourth sample that was annealed at 750 $^\circ$C for 3 h. This latter sample was also employed to study the effect of the misalignment between the membrane and substrate on the interface reconstruction, which in this case was estimated to be $\sim$4$^\circ$. SI Figure \ref{figimpact_mistilt} shows image series frames when applying our ``standard'' EDXS conditions of 250 pA beam current (electron flux $\sim$2.4 $\times$ 10$^{10}$ e$^-$~\AA$^{-2}$ s$^{-1}$). Even with this larger misalignment, crystal structure fills the $\sim$0.9 nm gap within the first 90 frames.
This observation implies that, even at larger misalignment angles, an ionic bond is formed between the membrane and substrate. With continued e-beam raster scanning, the misalignment between the membrane and the substrate is also corrected. However, in this case, only the misalignment of the first 3 unit cells of the substrate was corrected, rather than the deeper propagation of alignment seen in Figure \ref{fig:beam_effect}c.

\section {Supporting Information Figures}

\renewcommand{\thefigure}{S1}
\begin{figure}[H]
\centering
  \includegraphics[scale=0.7]{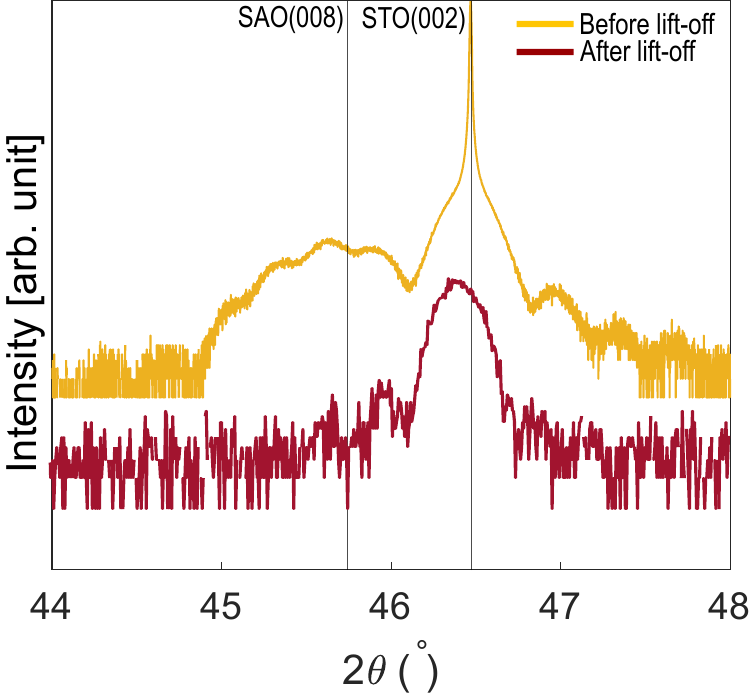}
  \caption{X-ray diffraction patterns of the SrTiO${_3}$(001)/Sr$_{3}$Al$_{2}$O$_{6}$/SrTiO${_3}$ heterostructure before lift-off (top curve), and of the SrTiO${_3}$ membrane on PDMS after lift-off (bottom curve). The patterns show that the crystalline structure of the SrTiO${_3}$ membrane is preserved after the lift-off procedure. Fitting the pattern of the lifted-off membrane on PDMS indicates a slight increase of the SrTiO${_3}$ out-of-plane \textit{c} lattice parameter, to a value of $\sim$3.91 \AA}
\label{fig:XRD}
\end{figure}

\renewcommand{\thefigure}{S2}
\begin{figure}[H]
\centering
  \includegraphics[scale=0.25]{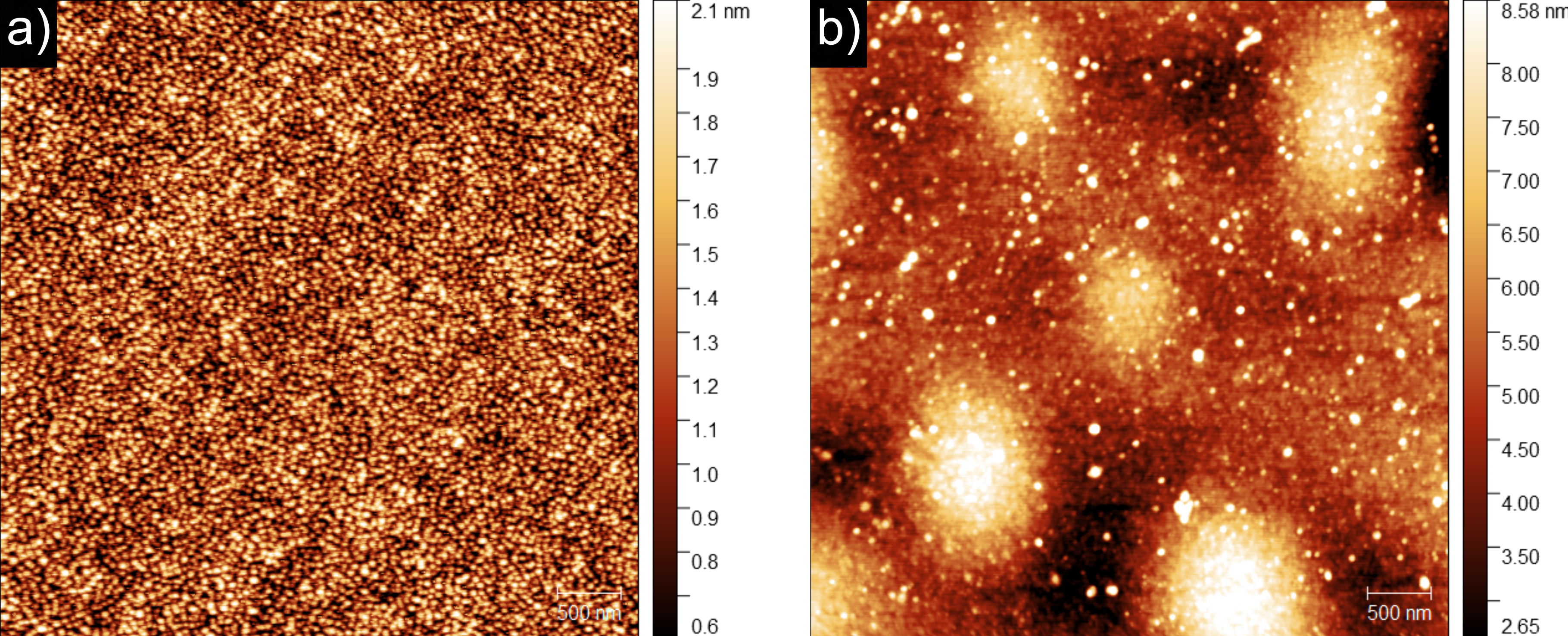}
  \caption{Surface topography acquired using atomic forces microscopy of: a) SrTiO${_3}$(001)/Sr$_{3}$Al$_{2}$O$_{6}$/SrTiO${_3}$ heterostructure before the lift-off procedure, with a root mean square (RMS) of $\sim$0.3 nm; b) Nb:SrTiO${_3}$(001)/SrTiO${_3}$ membrane after the transfer onto the Nb:SrTiO${_3}$(001) substrate, with a RSM of $\sim$1.3 nm.}
\label{fig:AFM}
\end{figure}

\renewcommand{\thefigure}{S3}
\begin{figure}[H]
\centering
  \includegraphics[width=\textwidth]{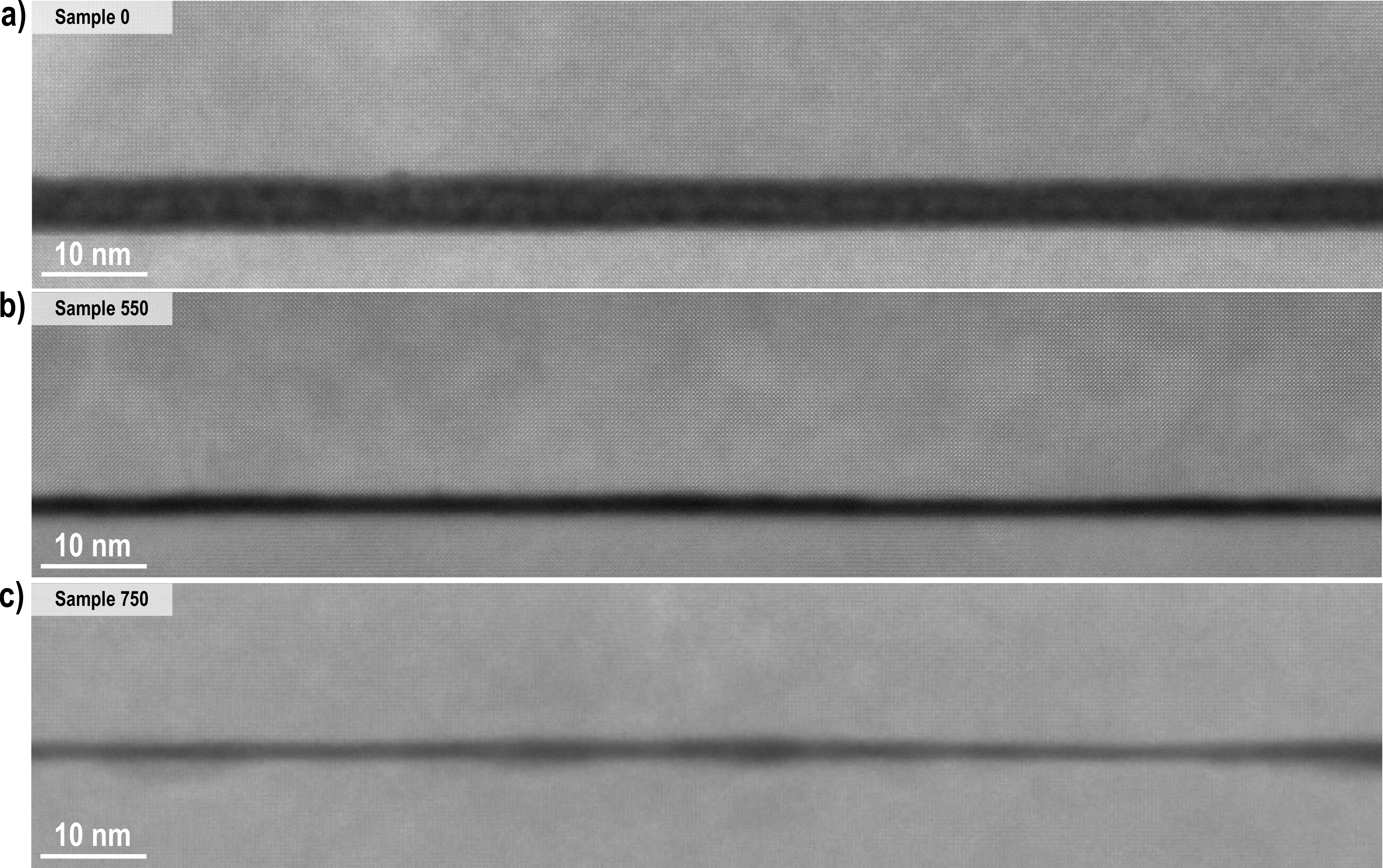}
  \caption{Large area cross-section HAADF STEM images of: a) Sample 0; b) Sample 550; c) Sample 750. A noticeable reduction of the interface gap width is observed after the thermal annealing.}
\label{fig:ADF_large_area}
\end{figure}

\renewcommand{\thefigure}{S4}
\begin{figure}[H]
\centering
  \includegraphics[width=\textwidth]{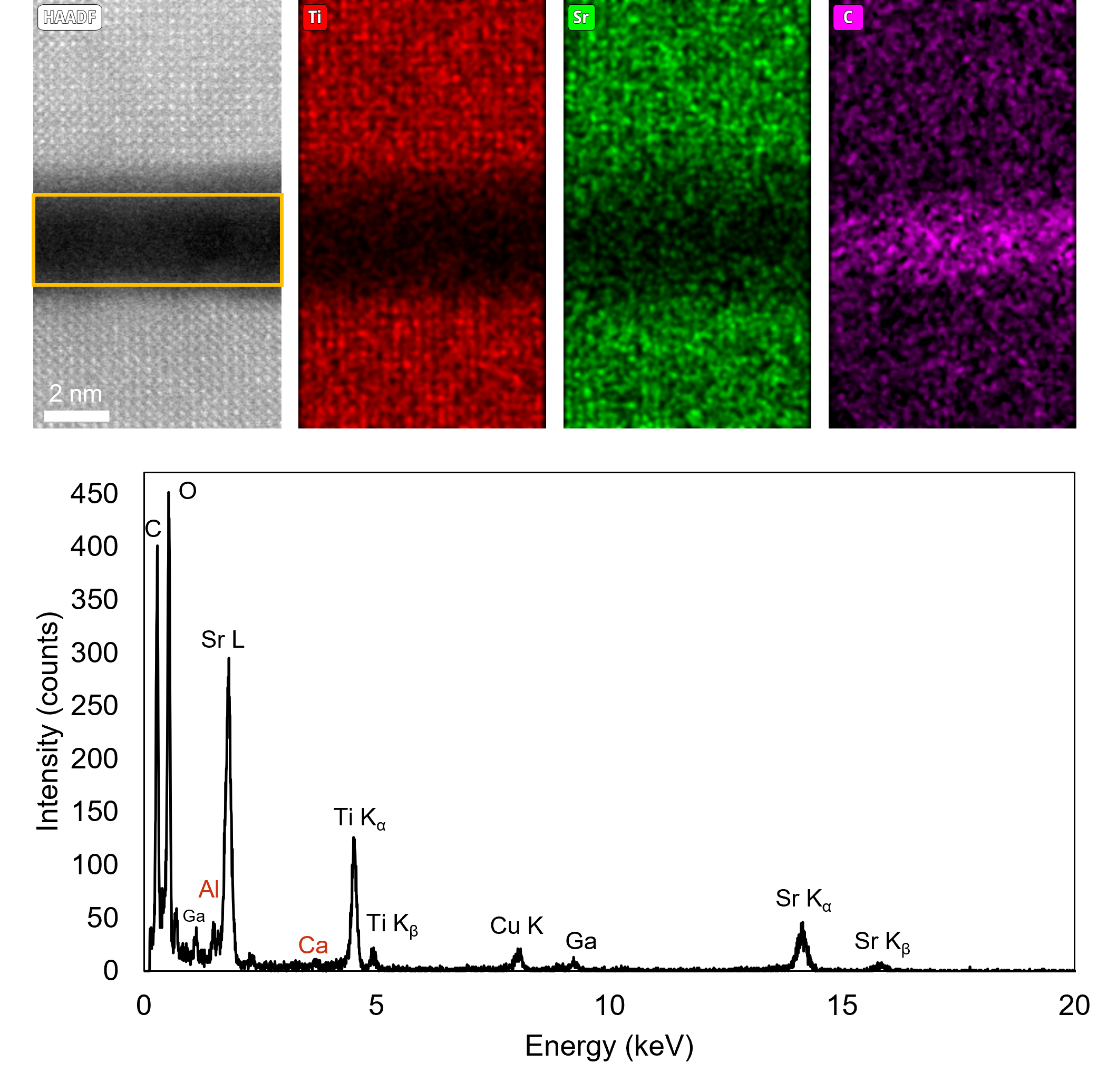}
  \caption{STEM-EDXS analysis of Sample 0 showing presence of Sr and Ti within the interface gap, along with some contaminant elements: C (contamination from contact with air); Al ($\sim$2 at.\% -- residual from Sr$_{3}$Al$_{2}$O$_{6}$); Ga (from focused ion beam milling sample preparation); and Ca ($\sim$1 at.\% -- residual from the lift-off process).}
\label{fig:EDX_sample0}
\end{figure}

\renewcommand{\thefigure}{S5}
\begin{figure}[H]
\centering
  \includegraphics[scale=0.65]{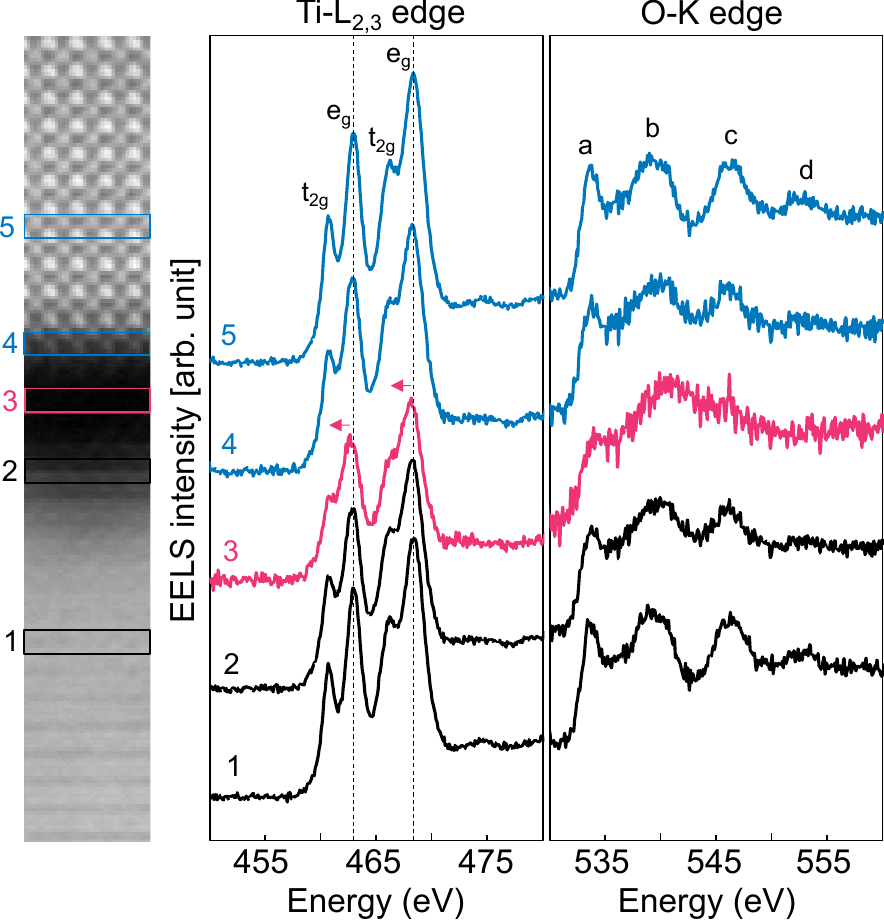}
  \caption{STEM-EELS analysis of Sample 550, from left to right: HAADF cross-section image of SrTiO${_3}$ membrane on Nb:SrTiO${_3}$(001) substrate after annealing at 550 $^\circ$C for 1 h; Ti-$L_{2,3}$ and O-$K$ edges extracted from \#1 Nb:SrTiO${_3}$(001) substrate, \#2 Nb:SrTiO${_3}$(001) substrate near the bottom interface, \#3 center of the gap, \#4 SrTiO${_3}$ membrane near the top interface, \#5 SrTiO${_3}$ membrane. The Ti-$L_{2,3}$ edge in spectrum \#3 shows some indication of splitting of the $L_3$ and $L_2$ peaks, with a slight left shift compared to the substrate spectra, revealing that the Ti begins to ``re-gain'' its $4+$ valence. Features also start to emerge in the O-$K$ edge, as compared to the as-transferred gap spectrum in Figure \ref{fig:annealing_effect}. Note that the HAADF image is cropped from the full width of the original mapped area.}
\label{fig:EELS_550}
\end{figure}

\renewcommand{\thefigure}{S6}
\begin{figure}[H]
\centering
  \includegraphics[width=\textwidth]{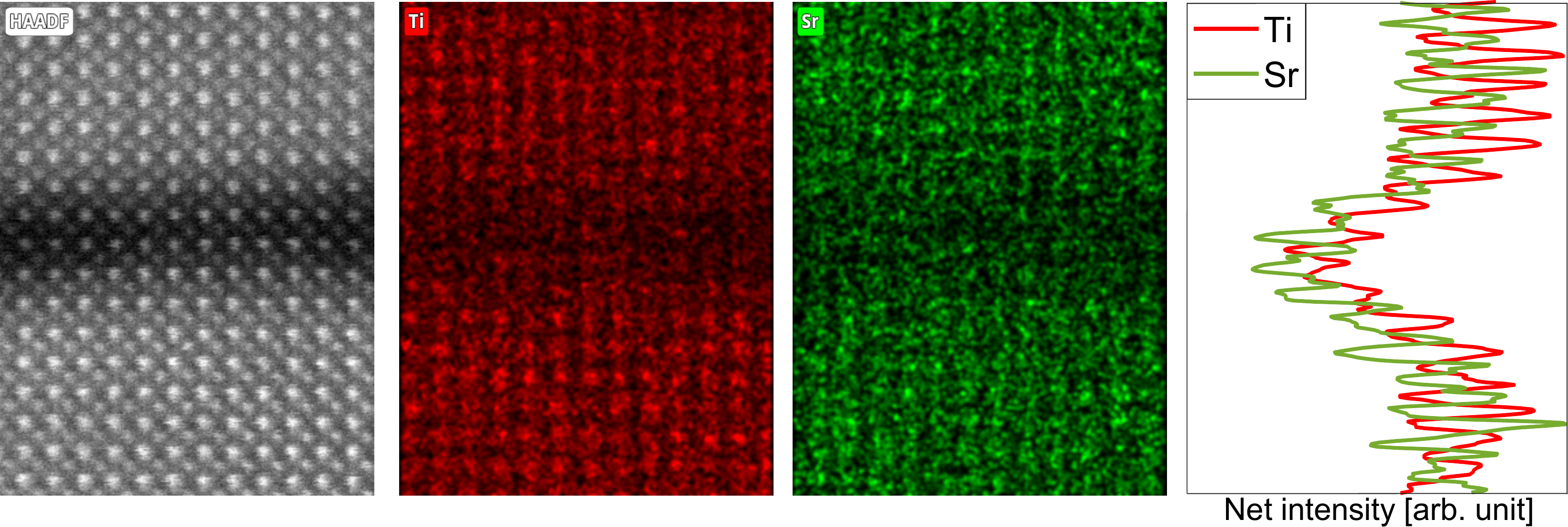}
  \caption{EDXS analysis of Sample 750 presented in Figure \ref{fig:beam_effect}c, integrating all spectral counts after 311 frames: final HAADF cross-section image of the reconstructed interface of Sample 750 (left); Sr and Ti net count maps (middle); d) integrated intensity line profiles of Sr and Ti across the reconstructed gap interface (right).}
\label{fig:haadf_profiles}
\end{figure}

\renewcommand{\thefigure}{S7}
\begin{figure}[H]
\centering
  \includegraphics[width=\textwidth]{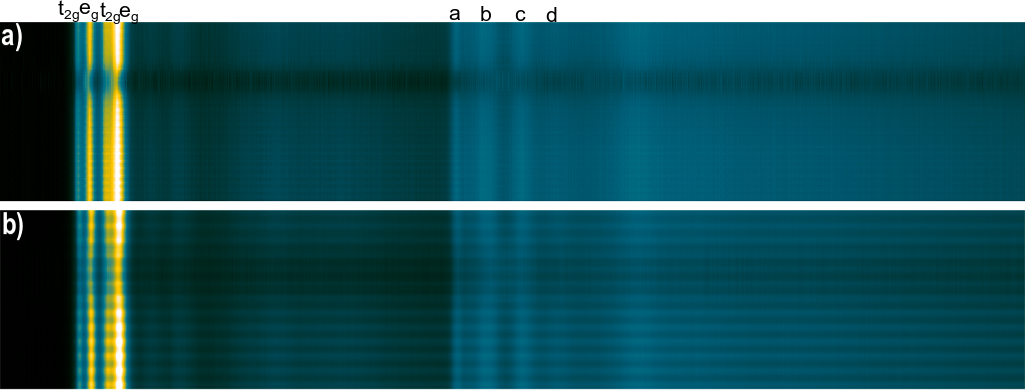}
  \caption{EEL spectra after background subtraction projected along a line in the out-of-plane direction, from before and after the local EDXS e-beam irradiation of Sample 750 that is discussed in Figure \ref{fig:comparison_annealing_beam_effect}: a) before EDXS acquisition; b) after EDXS acquisition. Both the increase in fine structure details of Ti-$L_{2,3}$ and O-$K$ edges and the much more defined spatial modulations in the gap from e-beam exposure are clearly discerned with this comparison.}
\label{fig:projected_edges}
\end{figure}

\renewcommand{\thefigure}{S8}
\begin{figure}[H]
\centering
  \includegraphics[width=\textwidth]{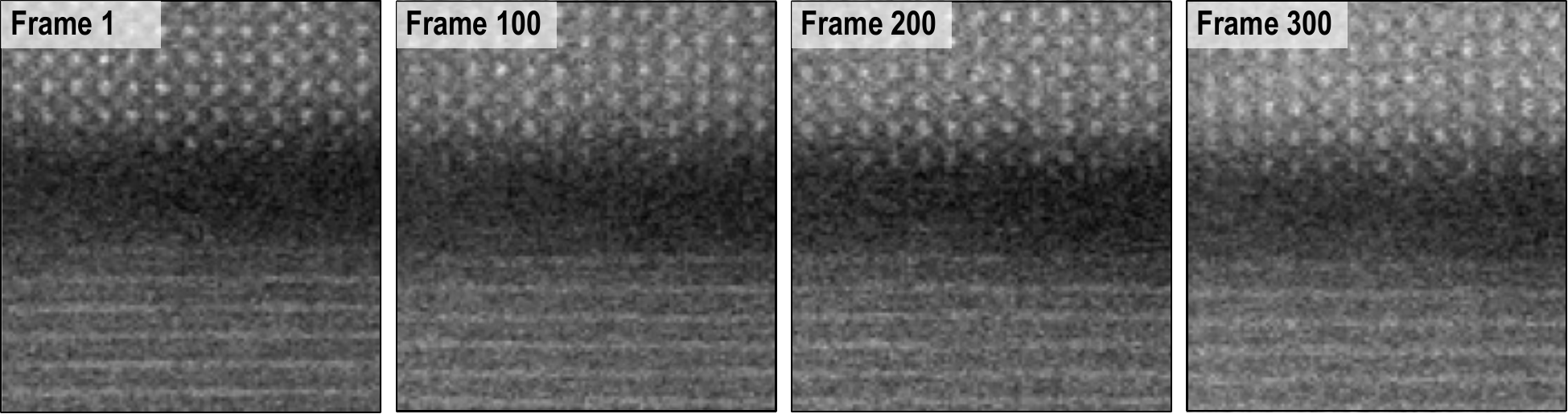}
  \caption{STEM HAADF frames from an EDXS acquisition of a fourth sample that was annealed at 750 $^\circ$C for 3 h, using a lower beam current of 100 pA, that corresponds to an electron flux of $\sim$10$^{10}$ e$^-$~\AA$^{-2}$ s$^{-1}$. No evolution of the nanostructure within the interface gap is observed between the frames. Not that the slight, cloudy increase in contrast by the last frame corresponds to typical contamination of the sample during the e-beam raster scanning.}
\label{fig:impact_current}
\end{figure}

\renewcommand{\thefigure}{S9}
\begin{figure}[H]
\centering
  \includegraphics[width=0.75\textwidth]{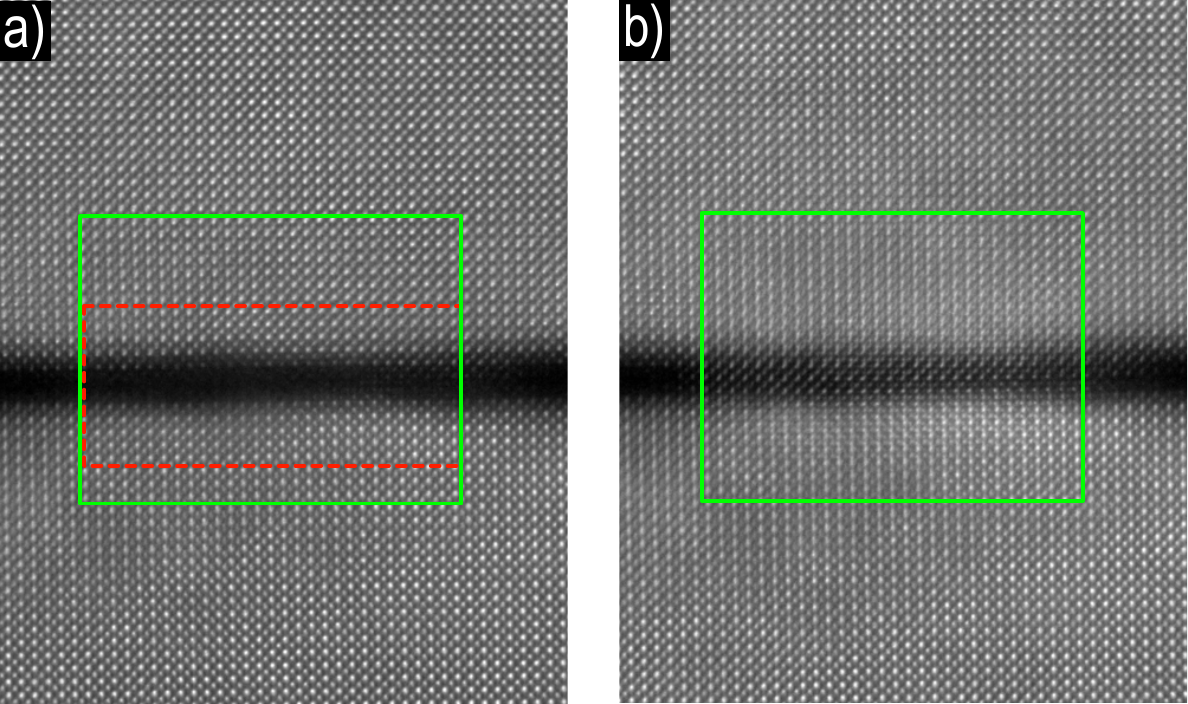}
  \caption{lower magnification HAADF STEM images of Sample 750: a) before local e-beam raster scan; b) after e-beam raster scan (using a ``focus window'' in DigitalMicrograph). The red box indicates the region that was raster scanned. From inspecting the HAADF image in Figure \ref{fig:focus_window}b, it is clear that the new crystalline structure in the interface gap is strongly localized to the raster-scanned region. Further, within the green box, the membrane and substrate crystal structures appear to twist into a new configuration that differs from the surrounding area not subjected to scanning, indicating the creation of local lattice distortions up to a depth of 7--10 unit cells into the pre-existing crystals.}
\label{fig:focus_window}
\end{figure}

\renewcommand{\thefigure}{S10}
\begin{figure}[H]
\centering
  \includegraphics[width=\textwidth]{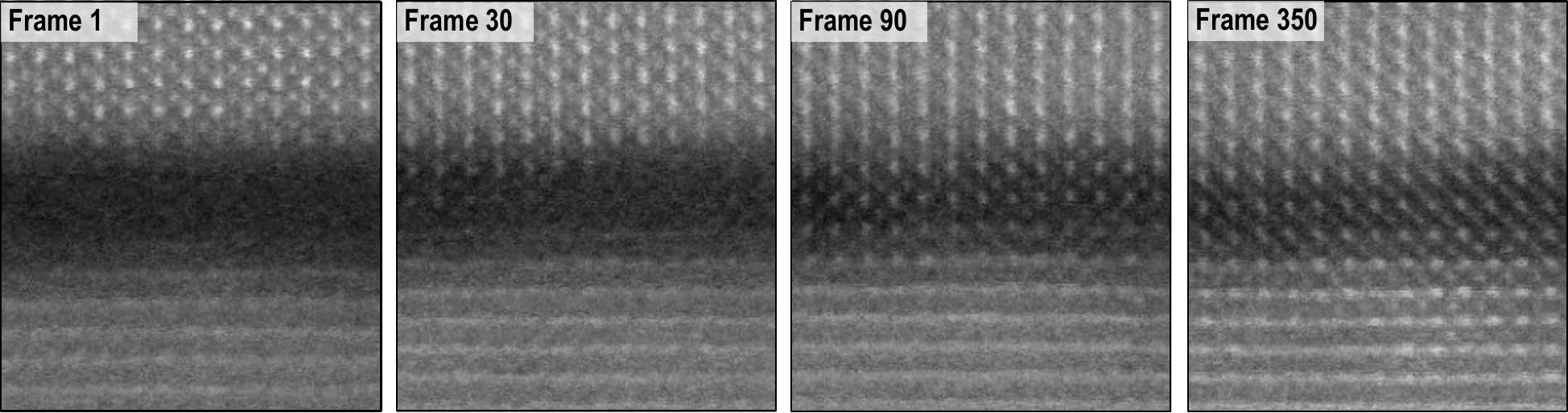}
  \caption{Impact of a larger misalignment between SrTiO${_3}$ membrane and Nb:SrTiO${_3}$(001) substrate on a fourth sample, annealed at 750 $^\circ$C for 3 h, having a misalignment between membrane and substrate of $\sim$4$^\circ$. Frames are shown from a local EDXS mapping acquisition using the standard e-beam current of 250 pA. By Frame 90, the $\sim$0.9 nm gap has been filled with new crystalline structure. By Frame 350, the substrate is showing structural realignment up to a depth of $\sim$3 unit cells.}
\label{figimpact_mistilt}
\end{figure}

\end{suppinfo}

\end{document}